\documentstyle[elsart12,rotate,epsf,openbib,shortcite]{elsart}
\def\ba{\begin{array}}
\def\ea{\end{array}}
\def\bc{\begin{center}}
\def\ec{\end{center}}
\def\be{\begin{equation}}
\def\ee{\end{equation}}
\def\bfe{\begin{figure}}
\def\efe{\end{figure}}
\def\btr{\begin{tabular}}
\def\etr{\end{tabular}}
\def\bea{\begin{eqnarray}}
\def\eea{\end{eqnarray}}
\def\beas{\begin{eqnarray*}}
\def\eeas{\end{eqnarray*}}

\def\astrobj#1{#1}

\def\arcsec{\hbox{$^{\prime\prime}$}}

\newcommand{\bj}{\rm B_{J}}
\newcommand{\bt}{\rm B_{T}\left(0\right)}

\newcommand{\dltwo}{\frac{\Delta L}{2}}
\newcommand{\dnb}{\delta\bar{n}}

\newcommand{\hnot}{\rm H_{o}}
\newcommand{\is}{I\left(R\right)\sigma_{\rm los}^{2}\left(R\right)}
\newcommand{\lll}{\log_{10}\left( L/L^{*}\right)}
\newcommand{\lmin}{L_{\rm min}\left( d_{i}\right) }
\newcommand{\lmax}{L_{\rm max}\left( d_{i}\right) }

\newcommand{\lv}{\log_{10}(\sigma}
\newcommand{\mstar}{\rm M_{B_{T}\left( 0\right) }^{*}}

\newcommand{\nvr}{\nu\overline{\sigma_{\rm r}^{2}}}
\newcommand{\phikp}{\phi_{k}^{\prime}}
\newcommand{\rc}{r_{\rm c}}
\newcommand{\re}{R_{\rm e}}
\newcommand{\resum}{\displaystyle\sum_{j=1}^{N_{\rm p}}\phi_{j}^{\prime}
\cdot Th\left[ L_{j}, \lmin, \lmax\right]}
\newcommand{\sr}{\overline{\sigma_{\rm r}^{2}}}
\newcommand{\st}{\overline{\sigma_{\theta}^{2}}}
\newcommand{\sphi}{\overline{\sigma_{\phi}^{2}}}
\newcommand{\thek}{Th\left[ L_{k}, \lmin, \lmax\right]}

\newcommand{\vdm}{\sigma_{\rm DM}}
\newcommand{\vff}{\sigma_{0.54\re}}
\newcommand{\vlos}{\sigma_{\rm los}\left( R\right)}
\newcommand{\vsre}{\sigma_{\re}^{*}}
\newcommand{\vvratio}{\sigma_{\rm los}^{2}/\vdm^{2}}

\journal{New Astronomy}
\begin{document}
\begin{frontmatter}

\title{An Elliptical Galaxy Luminosity Function and Velocity Dispersion Sample
of Relevance for Gravitational Lensing Statistics}

\author{Yu-Chung N. Cheng\thanksref{ycc}}
\author{and Lawrence M. Krauss\thanksref{lmk}}
\thanks[ycc]{Also in The MRI Institute for Biomedical Research,
E-mail address: yxc16@po.cwru.edu}
\thanks[lmk]{Also in Department of Astronomy, E-mail address:
krauss@theory1.phys.cwru.edu}

\address{Department of Physics, 
Case Western Reserve University, 10900 Euclid Ave.,
Cleveland, OH 44106-7079, USA}

\date{CWRU-P28-99}

\begin{abstract}
We have selected 42 elliptical galaxies from the literature and estimated their 
velocity dispersions at the effective radius ($\sigma_{\re}$) and at 0.54
effective radii ($\vff$). We find by a dynamical analysis that 
the normalized velocity dispersion of the dark halo of an elliptical galaxy 
$\vdm$ is roughly $\sigma_{\re}$ multiplied by
a constant, which is almost independent of the core radius or the anisotropy
parameter of each galaxy. Our sample analysis suggests that $\vdm^{*}$ 
lies in the range 
178-198 km s$^{-1}$. The power law relation we find between the 
luminosity and the dark matter velocity dispersion measured in this way is
$(L/L^{*}) = (\vdm/\vdm^{*})^\gamma$,  where $\gamma$ is between 2-3. These
results are of interest for strong gravitational lensing statistics studies.

In order to determine the value of $\vdm^{*}$, we calculate $\mstar$ in the
same $\bt$ band in which $\vdm^{*}$ has been estimated. We select 131 elliptical
galaxies as a complete sample set with apparent magnitudes $\bt$ between 9.26 
and 12.19. We find
that the luminosity function is well fitted to the Schechter form,
with parameters $\mstar$ = -19.66 + 5$\cdot\log_{10}h \pm 0.30$, $\alpha
= 0.15 \pm 0.55$, and the normalization constant $\phi^{*} = (1.34 \pm 0.30)
\times 10^{-3} h^{3}$ Mpc$^{-3}$, with the Hubble constant $\hnot$ = 100 $h$
km s$^{-1}$ Mpc$^{-1}$. This normalization implies that morphology type
E galaxies make up (10.8~$\pm$~1.2) per cent of all
galaxies.
\end{abstract}

\end{frontmatter}

\section{INTRODUCTION}

The search for the dark matter plays a central role in modern astrophysics
and cosmology. A key factor in constraining the 
cosmic abundance of dark matter
is to measure rotation curves or velocity dispersions of
galaxies. Because spiral galaxy rotation curves provide 
sensitive constraints on dark
matter, it is common to measure velocity dispersions in these systems. 
However, there are fewer available data sets of velocity dispersions for
elliptical galaxies.  

At the same time, the characteristic velocity dispersion of elliptical galaxies
is a very important parameter in the statistical study of strong gravitational
lensing, because the optical depth for lensing by galaxies is proportional to
the fourth power of the velocity dispersion \cite{TOGfirst}.   The value of the
normalized dark matter velocity dispersion for luminous elliptical galaxies 
that should be used in the strong gravitational lensing analysis is still under
some debate. For example, in 1984 Turner et al. \shortcite{TOGfirst} 
suggested  that one should use directly observed velocity dispersions, and
multiply them by
$\sqrt{3/2}$ to determine dark matter velocity dispersions, as suggested by the
virial theorem.  However, Kochanek 
\shortcite{Kochanek93,Kochanek94} has argued
using dynamical models that one should simply use the measured velocity
dispersions as the dark matter velocity dispersions for the lensing
calculation, without any
$\sqrt{3/2}$ correction.  However, even in this case, one must still 
estimate the characteristic dark matter velocity dispersion for ellipticals,
and this depends on one's assumptions about the elliptical galaxy luminosity
function \cite{chengkrauss}.   Moreover, both Kochanek
\shortcite{Kochanek93,Kochanek94} and  Turner et al. \shortcite{TOGfirst} focussed on
the singular  isothermal sphere (SIS) model in their gravitational
lensing analysis.  This model is clearly a very rough approximation to actual
galaxies.  

 In this paper, our goal is to re-examine 
both the appropriate luminosity function and velocity
dispersions for use in lensing statistics analyses.  In our analysis we
consider a finite core model, and consider the 
self-consistent determination of the velocity dispersion and luminosity
function for ellipticals.  This analysis is relevant even if the core
radius is quite small, or zero.
To be consistent we calculate the 
luminosity
function in the same band in which we estimate velocity dispersions.

We choose the Hubble constant $\hnot$ = 100 $h$ km s$^{-1}$ Mpc$^{-1}$ 
throughout.  We investigate the luminosity function 
in Section~\ref{sec:LF}. In section 3 we review our investigation of 
the theoretical determination of the dark matter velocity dispersion
in galaxies with core radii using  dynamical models with luminosity profiles
discussed by Jaffe \shortcite{Jaffe} and Hernquist \shortcite{Hernquist}.
 In Sections~\ref{sec:Re} 
and~\ref{sec:test}, we will 
describe our main results appropriate for use in lensing studies, 
including normalized 
velocity dispersions and their power law relations with the luminosity.

\section{THE LUMINOSITY FUNCTION}
\label{sec:LF}

The appropriate value of $\rm M^{*}$ in a Schechter formulation
\shortcite{Schechter} of the luminosity
 function will depend on which magnitude system one is measuring, so
that there is no unique value of
$\rm M^{*}$. In this paper, we will investigate the shape of the Schechter
luminosity function for a sample of  elliptical
galaxies in the $\bt$ band, as defined in de Vaucouleurs et al.
\shortcite{RC3} (hereafter RC3).
In order to calculate the luminosity of each galaxy in the $\bt$
band, we choose the distance of each galaxy from 
Faber et al. \shortcite{7s} (hereafter 7s),
and the apparent magnitude $\bt$ of each galaxy from RC3. We find 
415 galaxies in this set. However, we 
need to obtain a complete magnitude limited
sample set to calculate the luminosity function \cite{Loveday}. 
Based on the discussion in 7s, the elliptical galaxies in 7s are
complete to apparent magnitude 11.89. However, the RC3
catalogue was published after the appearance of 7s. Therefore, we consider 
galaxies with the apparent magnitudes given in 7s less than or equal to 11.89, 
and then check out their corresponding apparent magnitudes in RC3. We find the 
dimmest (or the largest number) apparent magnitude among these galaxies is 
12.19. Thus,
we use 12.19 as the cutoff apparent magnitude in RC3, and select those 
galaxies in 7s whose corresponding apparent magnitudes in RC3 are less than 
12.19. There are 131 galaxies in the final sample set.  We then determined
the luminosity function of this sample as described below.

\subsection{STY parametric maximum likelihood method}
\label{sec:STY}

We use here STY maximum likelihood method 
\cite{STYfirst}. There are some advantages to use this method,
as discussed in Efstathiou, Ellis \& Peterson \shortcite{EEP} 
(hereafter EEP). Consider the following probability
\be
p_{i} \propto \frac{\phi\left( L_{i}\right) }{\int_{\lmin}^{\lmax}
\phi\left( L\right) dL}
\label{eq:prob}
\ee
with Schechter \shortcite{Schechter} function
\be
\phi\left( L_{i}\right) = \frac{1}{L^{*}}\phi^{*}
\left(\frac{L_{i}}{L^{*}} \right)^{\alpha} 
\exp\left(-\frac{L_{i}}{L^{*}}\right)
\label{eq:sche}
\ee
This is the probability to find a galaxy at distance $d_{i}$ with luminosity
$L_{i}$ in the magnitude-limited sample set. $\lmax$ and $\lmin$
are calculated by using the distance $d_{i}$ (in unit of $h^{-1}$ Mpc) and 
the brightest and the dimmest
apparent magnitudes among the whole magnitude-limited sample set. For example,
the brightest apparent magnitude in our sample set is 9.26, so $\lmax/L^{*}$ 
will be equal to 
$d_{i}^{2}\cdot 10^{10 + 0.4 \left( {\rm M^*} - m_{\rm max} \right) }$, 
with $m_{\rm max}$ = 9.26. The likelihood function is $\cal L =
{\displaystyle\prod^N_{\rm i=1}}$$p_{i}$, and $\cal N$ is the total number of
galaxies in the sample set.

We maximize the likelihood function $\cal L$ with two parameters $\alpha$ and
$L^{*}$ (or $\rm M^{*}$). We do not additionally correct 
for a Malmquist bias because the
average dispersion between different $\rm B_T$ magnitudes in RC3 is small, and
also because Faber et al. \shortcite{7s} already corrected for the Malmquist 
bias
when they estimated the final distances of the galaxies in 7s. From the
maximum likelihood analysis, we get the parameters $\mstar$ = -19.66 + 5$\cdot
\log_{10}h \ {\displaystyle _{-0.33}^{+0.27}}$ and $\alpha$ = 0.15
${\displaystyle _{-0.54}^{+0.57}}$ (1 $\sigma$ error) (Figure~\ref{fig:STY}).

\bfe
{\hskip 4mm
\epsfxsize=80mm
\epsffile{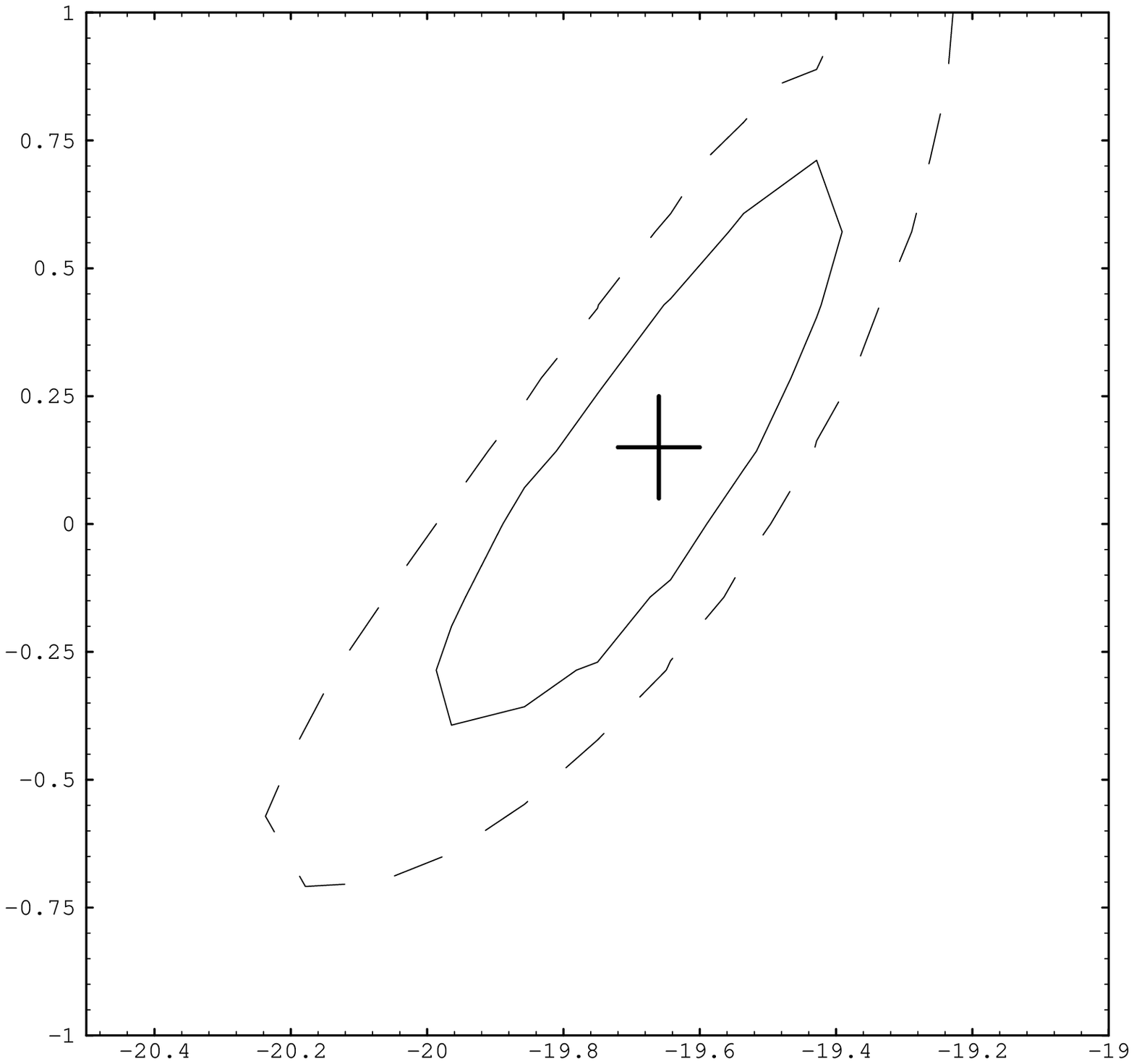}
}
\vskip -41mm
{\hskip 0mm $\alpha$}
\vskip 39mm
{\hskip 35mm ${\mstar} - 5\cdot\log_{10}h$} 
\vskip 2mm
\caption
{Parameters, $\alpha$ and ${\mstar} - 5\cdot\log_{10}h$, 
are fitted to the Schechter function by using STY maximum likelihood
method. The best fit has been marked `+.' The solid curve is the 1 $\sigma$
error contour, and the dash curve is the 95 per cent confidence level contour. 
The statistical method is discussed in the text.}
\label{fig:STY}
\efe

Because of the use of two different samples to attempt to obtain a complete
magnitude limited survey, one might be concerned that different choices of 
magnitude cutoff would produce different maximum likelihood estimates for the
luminosity function parameters.  
We explored a variety of combinations in
this regard, and found consistent results within the uncertainties quoted 
above.
For example, if we consider 
galaxies whose apparent magnitudes are below the apparent magnitude threshold
in 7s, then find the corresponding apparent magnitudes in RC3, and
neglect one apparently anomalous faint galaxy in 7s with apparent
magnitude 11.00 in RC3, then the next galaxy which is below the threshold
in 7s has  apparent magnitude 11.83 in RC3. If we use 11.82 as the cutoff
apparent magnitude, then we will have 82 galaxies in the sample set, and
then we get
$\mstar$ = -19.78
and $\alpha$ = 0.22. Choosing a fainter cutoff, 11.98, we find
$\mstar$ = -19.74 and $\alpha$ = 0.00 with 107 galaxies in the sample set.
As claimed, these are within the uncertainties in our quoted value above.

\subsection{Step-wise maximum likelihood method}
\label{sec:SWML}

In order to examine the goodness of fit of the Schechter function to 
the data we utilize a Step-wise maximum likelihood method to fit the
data (EEP).  This method is basically to replace the integral in
equation~(\ref{eq:prob}) with finite sum. Following presentation in
EEP, we define
\be
\phikp = \frac{\displaystyle\sum_{i=1}^{N} De\left( L_{i} - L_{k}
\right) }{\displaystyle\sum_{i=1}^{N} \frac{\thek}{\resum}}
\label{eq:swml}
\ee
where
\beas
\begin{array}{cl}
\phikp\left( L\right) & \equiv \phi^{\prime}\left( L_{k}\right)
{\hskip 0.5in \rm for}\; \left| L - L_{k} \right| < \dltwo\\

De(L_{i} - L_{k}) & =       \left\{ \begin{array}{cl}
1 & \mbox{if}\; \left| L_{i} - L_{k} \right| \leq \dltwo\\
0 & \mbox{others}
                                \end{array}
                        \right.\\

\thek &  =     \left\{{\hskip -0pt} \begin{array}{cl}
0 & {\rm if}\;\left\{{\hskip -0pt} \begin{array}{ll}
        L_{k}  <  \lmin - \dltwo \\
        L_{k}  >  \lmax + \dltwo
        \end{array}\right.\\
\frac{L_{k} - \lmin}{\Delta L} + \frac{1}{2} & {\rm if}\; \left| L_{k} -
\lmin\right| \leq \dltwo\\
\frac{L_{k} - \lmax}{\Delta L} + \frac{1}{2} & {\rm if}\; \left| L_{k} -
\lmax\right| \leq \dltwo\\
1 & {\rm others}
                                \end{array}
                        \right.
\end{array}
\eeas

\bfe
\bc\
\epsfxsize=15.0cm
\epsffile{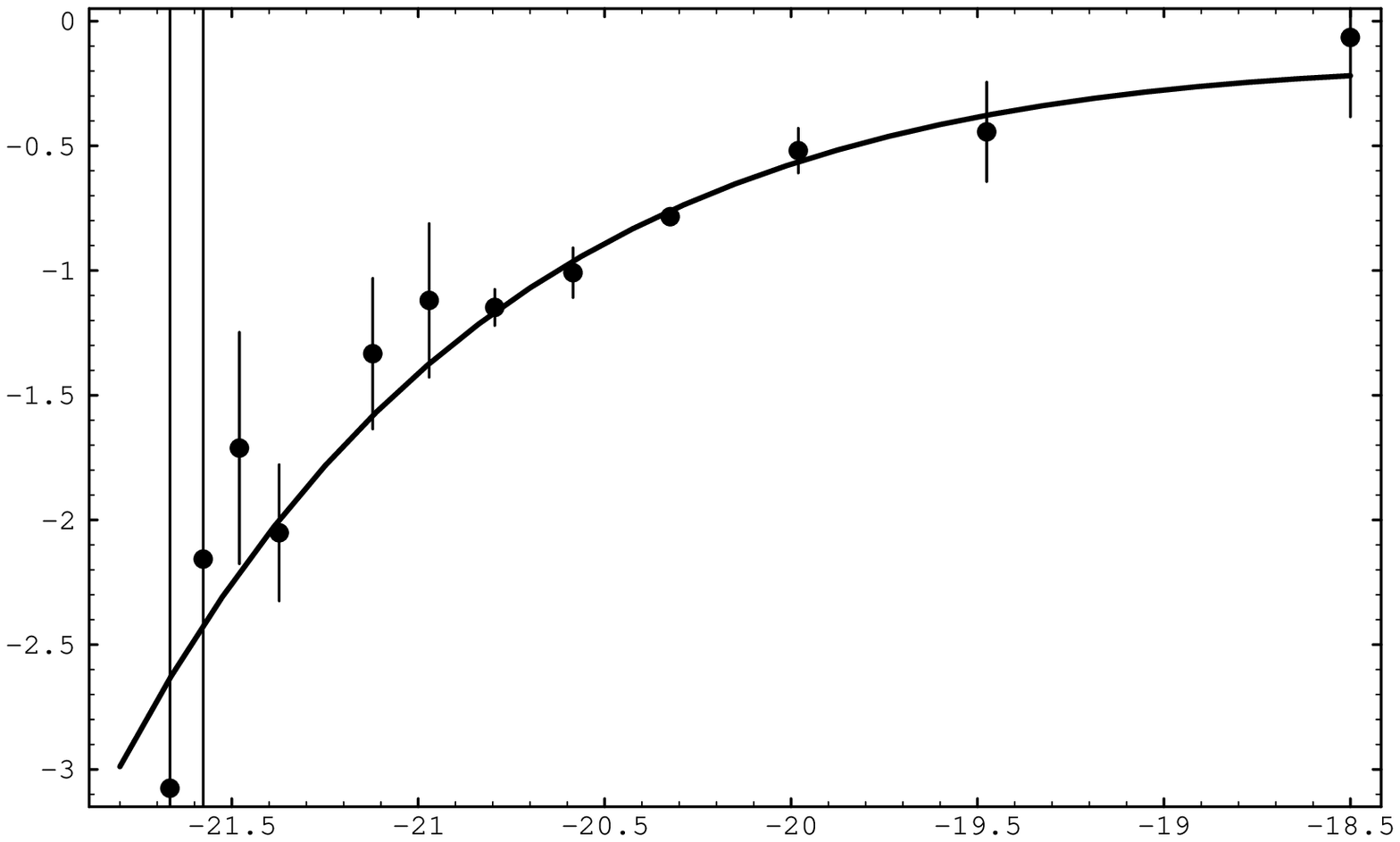}
\ec
\vskip -2.3in
\rotate[l]{\large $\log_{10}\left( \phi_{k}^{\prime}\right)$}
\vskip 0.3in
{\hskip 3.2in \Large bin width $\frac{\Delta L}{L^{*}}$ = 0.5}
\vskip 1in
\bc
{\large ${\rm M_{B_{T}(0)}} - 5\cdot\log_{10}h$}
\ec
\caption
{Comparison between the step-wise maximum likelihood method
(discrete dots) and the STY method (solid curve). We use results
from Section~\protect\ref{sec:STY}, and bin width $\Delta L/L^{*} = 0.5$ for 
the step-wise method in this plot.}
\label{fig:step}
\efe

Here $N$ is the total number of galaxies in the magnitude-limited sample
set, and
$N_{\rm p}$ is the total number of bins with bin width $\Delta L$. The index $k$
runs from 1, 2, 3, ......to $N_{\rm p}$. Here the step
function $Th[L_{k}, \lmin, \lmax ]$, is different from that in EEP,
because the
upper limit of the integral in equation~(\ref{eq:prob}) is $\lmax$ rather than
infinity. We simply use the results obtained in the previous subsection, put them
in
equation~(\ref{eq:swml}) to compare the values (along with errors) for
the discrete points derived in the second method to the Schechter
function form (Figure~\ref{fig:step}). We have to take some care to
normalize the values of 
$\phi_{k}^{\prime}$, if we want to compare these two maximum likelihood
fits.  
The uncertainty of each dot
is obtained using equation~(\ref{eq:swml}), using straightforward
error propagation \cite{Robinson}.

We can see from the figure that these two methods
are consistent with each other for the Schechter parameters from the
previous subsection, implying that the Schechter function is a good fit
to the data. This remained true for other choices of bin widths,
$\Delta L/L^{*}$ = 1 and $\Delta L/L^{*}$ = 0.1 as well. 

\subsection{The normalization constant}
\label{sec:norm}

We calculate the normalization constant $\phi^{*}$ using standard techniques
(EEP or Loveday et al. \shortcite{Loveday}). This involves calculating
the  average space density of galaxies in the luminosity
range being probed, $\bar{n}$, as
follows:
\be
\bar{n}  =  \frac{\displaystyle\sum_{i=1}^{N} w\left( d_{i}\right)}
{\int_{d_{\rm min}}^{d_{\rm max}}dx 4\pi x^{2} S\left(x\right) w\left( x\right)}
\label{eq:nbar}
\ee
where
\beas
S\left( d_{i}\right) & = & \frac{\int_{max\left( \lmin, L_{1}\right)}
^{min\left( \lmax, L_{2}\right)} \phi\left( L\right) dL}
{\int_{L_{1}}^{L_{2}} \phi\left( L\right) dL}\\
w\left( d_{i}\right) & = & \frac{1}{1 + 4\pi \bar{n}J_{3}\left(r_{\rm c}\right)
S\left( d_{i}\right)}\ , {\hskip 0.5in}
J_{3}\left( r_{\rm c}\right) = \int_{0}^{r_{\rm c}} r^{2} \xi\left( r\right) dr
\eeas 
$N$ is the total number of galaxies in the sample set, $d_{i}$ is the distance 
of each galaxy, and $d_{\rm min}$ and $d_{\rm max}$ are the nearest and the 
furthest distances among galaxies in the sample set. We have $d_{\rm min}$ = 
6.95 $h^{-1}$ Mpc, and $d_{\rm max}$
= 57.79 $h^{-1}$ Mpc. $L_{1}$ and $L_{2}$ are related to the selection
function $S(d_{i})$ and the weighting function $w(d_{i})$, but these can,
with sufficient accuracy for our purposes be chosen to be
$\lmin$ and $\lmax$, the dimmest and the brightest luminosities from our
sample set. Finally, $\xi(r)$ is the two-point galaxy correlation
function with $r_{\rm c} \approx 20 h^{-1}$ Mpc, and we use $4\pi J_{3} \approx
10,000 h^{-3}$~Mpc$^{3}$ here \cite{EEP,Loveday}.

After applying the Schechter function in the selection function $S(d_{i})$, we
find $\bar{n}$ = 9.65 $\times$ 10$^{-4} h^{3}$ Mpc$^{-3}$ from 
equation~(\ref{eq:nbar}). The normalization constant is equal to
\be
\phi^{*} = \frac{\bar{n}}{\int_{L_{1}}^{L_{2}} \phi\left( L\right) dL}
\label{eq:norm}
\ee
So we get $\phi^{*}$ = 1.34 $\times$ 10$^{-3} h^{3}$ Mpc$^{-3}$.
The uncertainty of $\bar{n}$ is approximated by \cite{Loveday}
\be
\dnb \approx \sqrt{\frac{\bar{n}}{\int_{d_{\rm min}}^{d_{\rm max}} wS dV}}
\ee
We thus find $\dnb/\bar{n} \approx$ 0.19. This value is larger than that in Loveday 
et al.
\shortcite{Loveday}, because $\dnb/\bar{n}$ is inversely proportional to
the square 
root of number of galaxies in the sample set. There are 131 galaxies in our 
sample set, but there were about 1600 galaxies in Loveday et al.
\shortcite{Loveday}. We thus find the uncertainty of $\phi^{*}$ by standard
methods of error propagation, and hence obtain our final result, 
$\phi^{*} = (1.34 \pm 0.30)
\times 10^{-3} h^{3}$ Mpc$^{-3}$.

\section{MODELS OF VELOCITY DISPERSION}
\label{sec:VD}

We have already mentioned the importance of 
using a self-consistent analysis of dark matter 
velocity dispersions when attempting
to model galaxies as lenses \cite{chengkrauss}.  
Here we describe such an analysis for various
galaxy models.  The key point is that the determination of $\vdm$ is
a sensitive function of the galaxy anisotropy parameters and core radius, and
one must consider this sensitivity when determining $\vdm$ on the basis of
various observations. A theoretical dynamical analysis suggests that measuring
line of sight velocity dispersions at the de Vaucouleurs effective radius 
$\re$ \shortcite{Vaucouleurs},
can provide a relatively robust estimate of the underlying $\vdm$, and
also allow one to extract out contributions due to central mass concentrations
in the galaxy.

\bfe
\epsfxsize=16.0cm
\epsffile{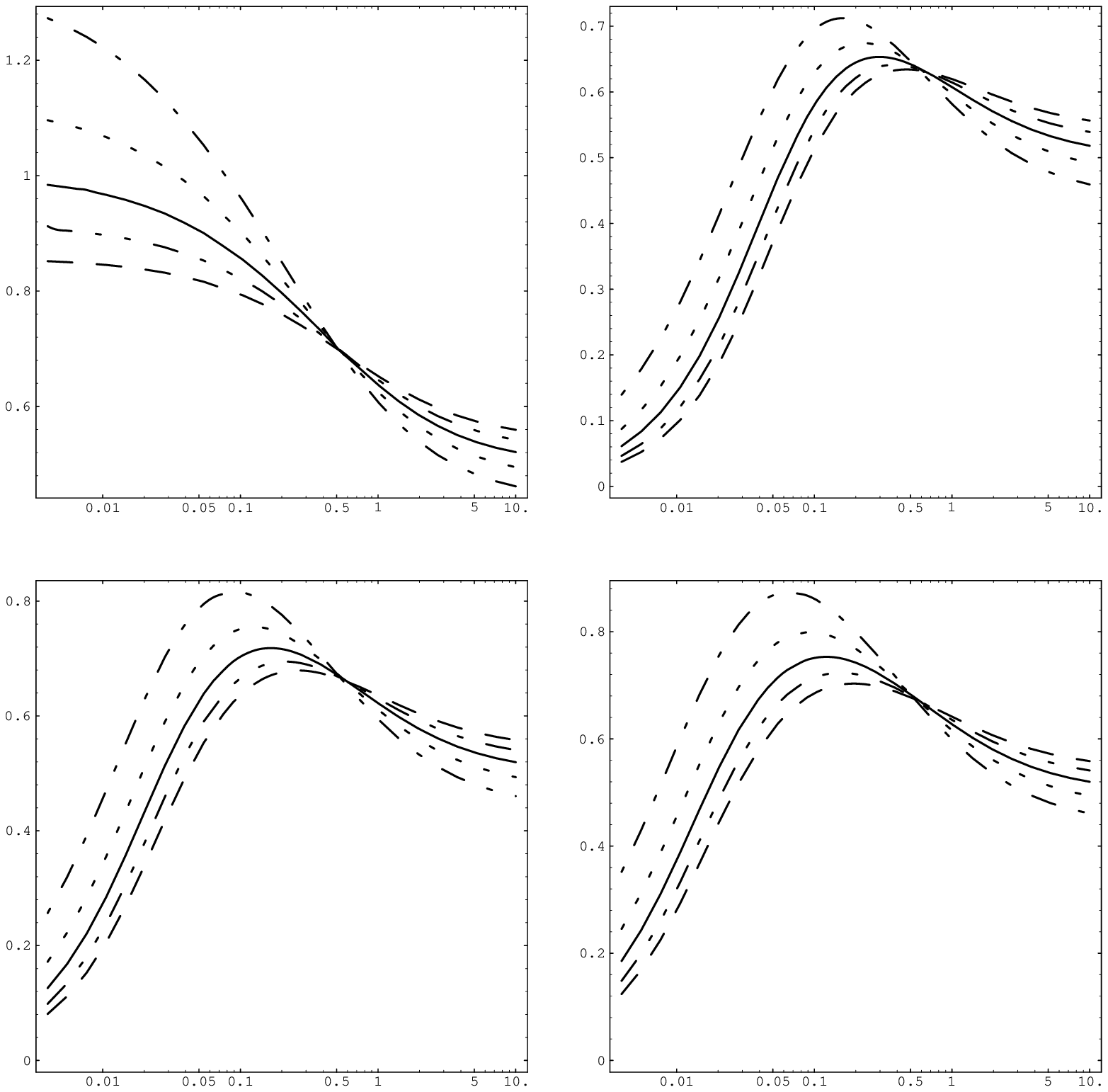}

\vskip -13.80cm
{\hskip 1.8in $\rc = 0$}
\vskip 10pt
{\hskip -12pt \rotate[l]{$\vvratio$}}
{\hskip 3.05in \rotate[l]{$\vvratio$}}
\vskip 1.10cm
{\hskip 4.70in $\rc = \re/60$}
\vskip 1.40cm
{\hskip 1.3in $R/\re$}
{\hskip 2.8in $R/\re$}
\vskip 3.10cm
{\hskip -12pt \rotate[l]{$\vvratio$}}
{\hskip 3.05in \rotate[l]{$\vvratio$}}
\vskip 0.50cm
{\hskip 1.8in $\rc =\re/40$}
{\hskip 2.1in $\rc = \re/20$}
\vskip 1.95cm
{\hskip 1.3in $R/\re$}
{\hskip 2.8in $R/\re$}
\vskip 0.1cm
\caption
{Jaffe~\protect\shortcite{Jaffe} luminosity profile with core radii, 0, $\re/20,
\re$/40, and $\re$/60. The five curves in each plot, dash-dot curve, dot-dot 
curve, solid
curve, dash-dot-dot curve, and dash-dash curve, represent the five different
anisotropy parameters $\beta$ = 0.4, 0.2, 0.0, -0.2, and -0.4, respectively.}
\label{fig:Jaffe}
\efe

\bfe
\epsfxsize=16.0cm
\epsffile{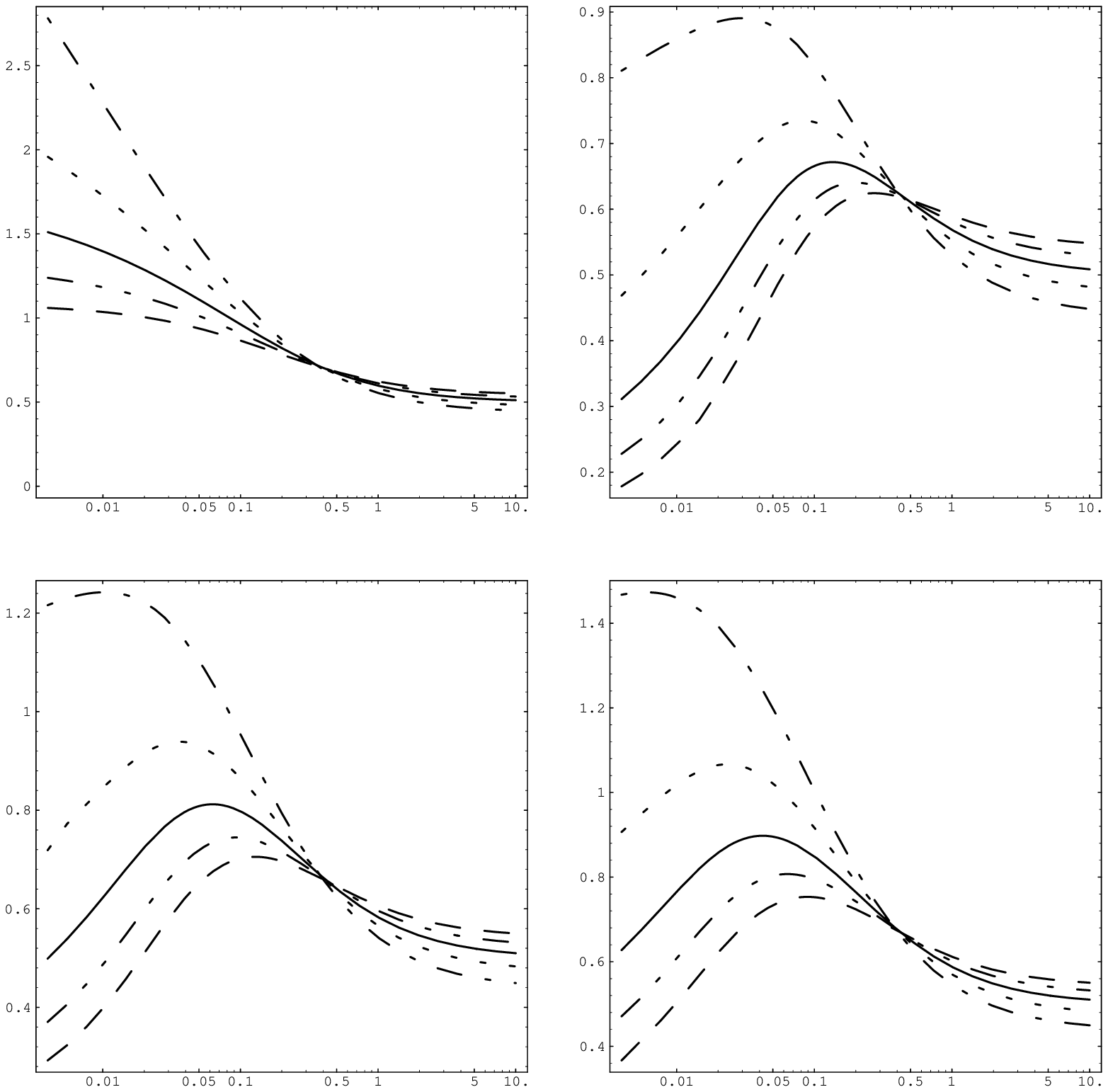}
\vskip -13.80cm
{\hskip 1.8in $\rc = 0$}
{\hskip 2.50in $\rc = \re/60$}
\vskip 10pt
{\hskip -12pt \rotate[l]{$\vvratio$}}
{\hskip 3.05in \rotate[l]{$\vvratio$}}
\vskip 2.50cm
\vskip 12pt
{\hskip 1.3in $R/\re$}
{\hskip 2.8in $R/\re$}
\vskip 1.70cm
{\hskip 1.8in $\rc =\re/40$}
\vskip 25pt
{\hskip -12pt \rotate[l]{$\vvratio$}}
{\hskip 3.05in \rotate[l]{$\vvratio$}}
\vskip 1.00cm
{\hskip 4.7in $\rc = \re/20$}
\vskip 1.50cm
{\hskip 1.3in $R/\re$}
{\hskip 2.8in $R/\re$}
\vskip 0.1cm
\caption
{Hernquist \protect\shortcite{Hernquist} luminosity profile with core radii,
0, $\re/20, \re$/40, and $\re$/60. The description of the five curves in each
plot is the same as in Figure~\protect\ref{fig:Jaffe}.}
\label{fig:Hernquist}
\efe

Throughout our analysis we model the
mass density distribution of elliptical galaxies with the
following form \cite{HK}:
\be
\rho = \frac{\vdm^{2}}{2\pi G\left( r^{2} + \rc^{2}\right)}
\label{eq:dens}
\ee
where $\vdm$ (independent of radius $r$) is the velocity dispersion of this 
system (presumably the dominant dark matter), and
$\rc$ is the core radius. When $\rc$ is zero, this model reverts to a singular
isothermal sphere (SIS) model. 
We should bear in mind that the
velocity
dispersion in this equation, which is assumed to be independent of radius,
cannot be measured directly.
What we can measure is the line-of-sight velocity dispersion
$\vlos$, with the projected distance $R$ measured from the center of the
observed galaxy (i.e., $R$ is perpendicular to the line-of-sight to us).
This means that we need to integrate over the line-of-sight distance in
order to work backward to infer the value of $\vdm$ for theoretical purposes.
For a singular isothermal sphere, $\vdm $ is not a function of $R$, but
$\vlos$ is, and this relationship is different when a galaxy has
a finite core (although this is a secondary issue).  Thus, one
has to be careful how to use the measured velocity dispersion to derive the
relevant quantity to utilize for lensing, namely whether it well approximates
$\vdm$.
To  find the relationship between $\vdm$ and $\vlos$, we start from Jeans
equations  as follows \cite{Ogorodnikov,BT}:
\be
\frac{d}{dr}\left( \nvr\right) + 2\beta\frac{\nvr}{r} = -\nu\frac{d\Phi}{dr}
\label{eq:jean}
\ee
where
\be
\frac{\partial\Phi}{\partial r} = \frac{4\pi G}{r^{2}}\int_{0}^{r}
dr^{\prime}\cdot {r^{\prime}}^{2}\rho\left( r^{\prime}\right) =
2\vdm^{2}\left[\frac{1}{r} - \frac{\rc}{r^{2}}\arctan\left(\frac{r}{\rc}
\right)\right]
\label{eq:poten}
\ee
by using equation~(\ref{eq:dens}).

In equation~(\ref{eq:jean}), $\nu(r)$ is the luminosity profile, 
$\sigma_{\rm r}\left(r\right)$ is the radial part of the velocity dispersion
of the observed galaxy at distance $r$ from the center of that galaxy. 
The other two components of the velocity dispersion $\sigma$ in spherical
coordinates are $\sigma_{\theta}$ and $\sigma_{\phi}$.
We define the anisotropy $\beta(r)$ as $\st = \sphi = (1 - \beta) \sr$. 

The line-of-sight velocity dispersion $\vlos$ is related to $\nvr$ by
\be
\is = 2\int_{R}^{\infty}\left( 1 - \beta\frac{R^{2}}{r^{2}}\right)\nvr
\frac{r}{\sqrt{r^{2} - R^{2}}}dr
\label{eq:ilos}
\ee
where $I(R)$ is the surface brightness profile, which can be calculated
as follows:
\be
I\left(R\right) = 2\int_{R}^{\infty}\frac{\nu r}{\sqrt{r^{2} - R^{2}}}dr
\label{eq:brit}
\ee

By solving the differential equation~(\ref{eq:jean}), we cam re-write
equation~(\ref{eq:ilos}) as 
\be
\is = 2\int_{0}^{\infty}du\left( 1 - \beta\frac{R^{2}}{R^{2} + u^{2}}\right)
\frac{1}{\left( R^{2} + u^{2}\right)^{\beta}} 
\int_{\sqrt{R^{2} + u^{2}}}^{\infty} 
dx\cdot x^{2\beta}\nu\left( x\right)\frac{\partial\Phi}{\partial x}
\label{eq:model}
\ee

If we assume $\beta$ is simply a constant independent of $r$ in equations~(\ref{eq:poten}),~(\ref{eq:brit}), and~(\ref{eq:model}), then once we know the
form for the luminosity profile $\nu(r)$, we know the ratio
of $\vvratio$ at the different projected distance $R$. 
We have considered two different but similar luminosity profiles
$\nu(r)$.  We adopt $\nu(r) \propto r^{-2}(r + r_{0})^{-2}$, with $r_{0}
= 1.31\re$, from Jaffe \shortcite{Jaffe}, and $\nu(r) \propto r^{-1}(r +
r_{0})^{-3}$,  with 
$r_{0} = 0.45\re$, from Hernquist \shortcite{Hernquist}. We performed our 
numerical calculations by adjusting
$\beta$ = 0.4, 0.2, 0, -0.2, -0.4 and $\rc$ = 0, $\frac{1}{20}\re, 
\frac{1}{40}\re, \frac{1}{60}\re$ for both Jaffe \shortcite{Jaffe} and 
Hernquist \shortcite{Hernquist} luminosity profiles (Figure~\ref{fig:Jaffe} and 
Figure~\ref{fig:Hernquist}). The ratio of the core radius to the
effective radius may be estimated from Lauer \shortcite{Lauer}, and we find
that this ratio is between 1/20 and 1/60 for class I galaxies in Lauer 
\shortcite{Lauer}. From Figure~\ref{fig:Jaffe} and Figure~\ref{fig:Hernquist},
we clearly see that $\vvratio$ is strongly sensitive to
the anisotropy
when the projected radius is less than 0.1$\re$ (whether or not the
core radius is non-zero, a factor to which the result is also sensitive).
This is consistent with observational data \cite{BSGfirst}.
It can thus be dangerous to simply
consider the line-of-sight velocity dispersion within 0.1$\re$ as $\vdm$.
 Even if the core radius
is effectively very small, as numerous observations of ellipticals suggest,
the sensitivity to the anisotropy parameter implies one should be
careful when interpreting meassured values of the velocity dispersion.

However, if we are careful to look at all the 40 curves in 
Figure~\ref{fig:Jaffe} and Figure~\ref{fig:Hernquist}, we can find the following
inequalities: 

$\ba{cccccl}
1.16 & \leq & \frac{\vdm}{\vlos} & \leq & 1.27 & {\hskip 0.2in} {\rm at}\; R = 0.4\re \\
1.20 & \leq & \frac{\vdm}{\vlos} & \leq & 1.30 & {\hskip 0.2in} {\rm at}\; R = 0.54\re\\
1.24 & \leq & \frac{\vdm}{\vlos} & \leq & 1.37 & {\hskip 0.2in} {\rm at}\; R = \re
\ea$\\

These indicate that the intrinsic scatter of $\vdm$ will be less 
than 10 per cent if we can
measure the line-of-sight velocity dispersion at $\re$ or half $\re$. Then, we
can multiply this velocity dispersion by the average value (e.g., 1.31 for $R$
= $\re$, or 1.25 for $R$ = $0.54\re$), in order to get $\vdm$. We can see that 
this argument is almost independent of the core radius 
of each galaxy and the anisotropy parameter $\beta$. We 
consider below the velocity dispersions at two different $R$ values as a 
consistency check of our analysis.  First however, we comment on another
important feature of these figures which is relevant to a determination
of galaxy parameters for lensing purposes.

Kochanek \shortcite{Kochanek96}
has pointed out that the inclusion of a core radius will in general
require, for self-consistency, a larger value of $\vdm$ to be used in 
models, and has argued that this will cancel out the effect that non-zero
core radii can have on suppressing lensing optical depths if the value of
$\vdm$ were instead kept fixed.  Examining Figures~\ref{fig:Jaffe}
and~\ref{fig:Hernquist} at approximately
0.5 $\re$, where the different curves intersect, indeed illustrates the
effect described by Kochanek. As the core radius is increased $\vdm$ 
increases.
However the key point is that the change in $\vdm$ is 
small compared to the huge
variances in $\vdm$ one sees in models with different values of $\beta$,
and also is small compared to the possibility of overestimating $\vdm$ if
central line-of-sight velocity dispersions are used.

Note that in this section we chose
42 galaxies from 12 different data sets to be our galaxy sample, because
for these we could find or estimate the velocity
dispersions at both $\re$ and $0.54\re$ . Among those 12
sets of observations, most use a slit size from one to three
arcseconds. This should therefore not be significant when
 $\re$ is on the  order of ten arcseconds, as it is in this
paper (see Table~\ref{table:Re}). Furthermore, from
Figures~\ref{fig:Jaffe} and~\ref{fig:Hernquist}, one can see that the
velocity curves are relatively smooth between 
$\re$ and $0.54\re$ (i.e., theoretically, $\vdm=1.31\sigma_{\re}=1.25\vff$).
This indicates that even if the slit size were up to half $\re$, this
effect might still not strongly bias the results.

\subsection{Velocity dispersions at $R = \re$}
\label{sec:Re}

Based on values of $\re$ from RC3, we can estimate 
velocity dispersions $\vdm$ of these 42 galaxies. 
We also consider $\re$ from various different papers and 
translate the different values of velocity dispersions into the uncertainty of
the velocity dispersion measured at $\re$, as given by RC3. We would 
quote either this uncertainty or the uncertainty from the observations, 
whichever is larger. The other problem is that different observers might
use different methods to measure velocity dispersions.
 We calculate the weighted mean of all these 
velocity dispersions for one galaxy as our final line-of-sight velocity 
dispersion of that galaxy \cite{Lauer,Robinson}. 
We list the information of $\re, \bt$, distances and velocity dispersions
of all 42 galaxies in Table~\ref{table:Re}. 

Using Table~\ref{table:Re} and a suitable value for 
$\mstar$ = -19.66 + 5$\log_{10}h \pm 0.30$ as discussed in Section~\ref{sec:LF},
we can calculate the least-square-fit of the
relation between $\lll$ and $\lv_{\re}$/km s$^{-1})$ (Figure~\ref{fig:Re}). We
exclude galaxies \astrobj{NGC 4251}, \astrobj{NGC 7796}, \astrobj{NGC 4486B}, 
and \astrobj{NGC 6411}, because the
former two galaxies do not have available distances from 7s, and the
latter two galaxies are extreme outlyers (One galaxy 
is the farthest right one and the other galaxy is the lowest one in 
Figure~\ref{fig:Re}.), and there is 
no available observational value of $\re$ of \astrobj{NGC 4486B} in RC3. 
We obtain
\be
\lll = (-4.04 \pm 0.49) + (1.89 \pm 0.22)\lv_{\re}/{\rm km\ s}^{-1})
\label{eq:vre}
\ee
with $\chi^{2}$ = 48 for 38 galaxies in Figure~\ref{fig:Re}. 
We use both uncertainties of luminosities and velocity dispersions to work
out uncertainties in equation~(\ref{eq:vre}) and its $\chi^{2}$. If we
set $L = L^{*}$ in equation~(\ref{eq:vre}), then we find the velocity
dispersion of luminous elliptical galaxies at the effective radius,
$\vsre = 136 \pm 15$ km s$^{-1}$. From the discussion in Section~\ref{sec:VD},
we then
multiply this number by 1.31, to get $\vdm^{*} \approx 1.31\vsre \approx$
178 km s$^{-1}$. 

\bfe
\bc\
\epsfxsize=15.0cm
\epsffile{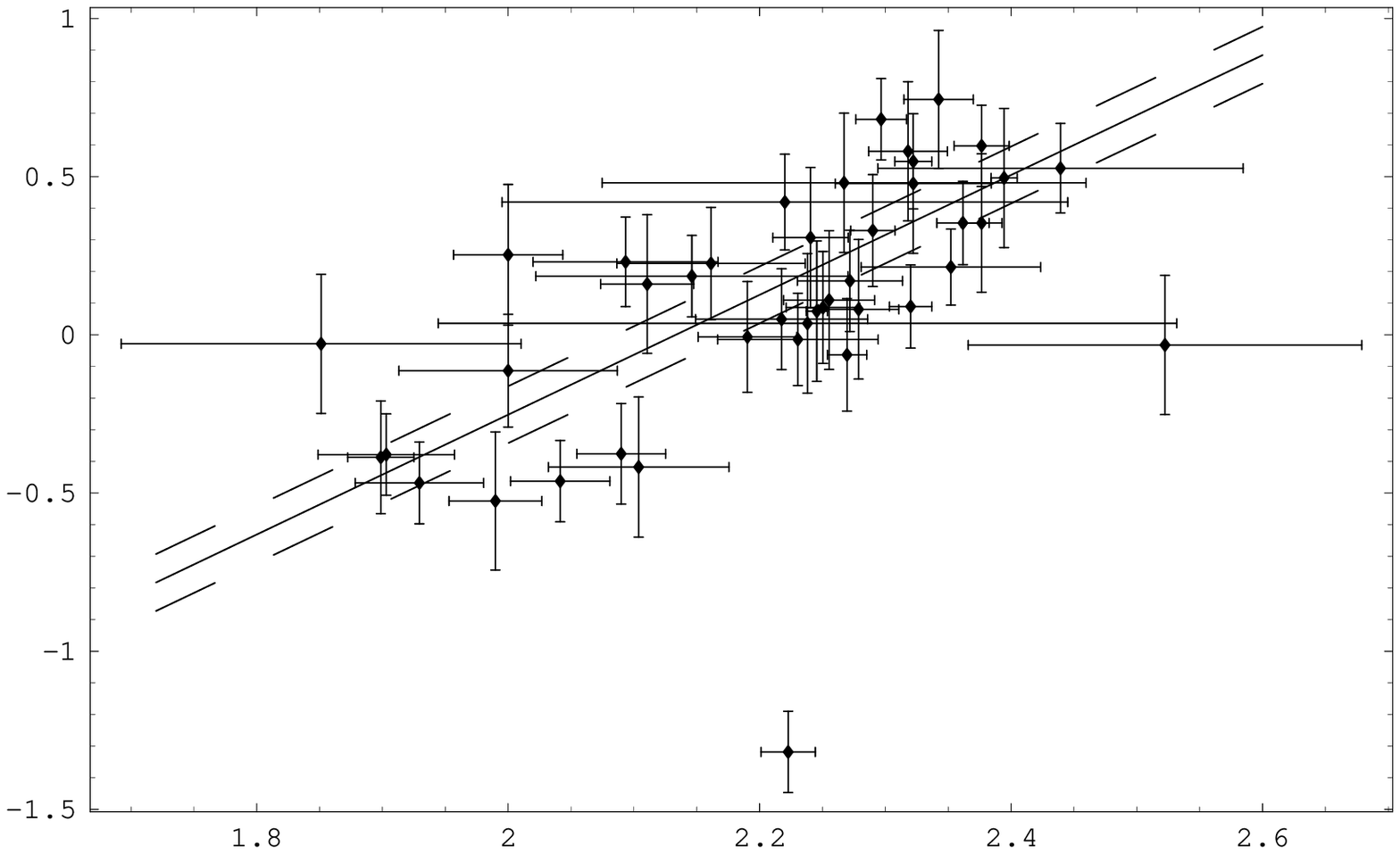}
\ec
\vskip -2.60in
\rotate[l]{\hskip -00pt \large $\lll$}
\vskip 0.3in
{\hskip 3.2in \Large $\chi^{2} = 48$}
\vskip 1.05in
\bc
{\large $\lv_{\re}$/km s$^{-1})$}
\ec
\caption
{Least-square-fit of $\lll$ vs. $\lv_{\re}$/km s$^{-1})$. The center solid line
is the best fit, as shown in equation~(\protect\ref{eq:vre}). The upper dash
line is the best fit plus 0.09, and the lower dash line is the
best fit minus 0.09. These two lines are the 95 per cent confidence limits.}
\label{fig:Re}
\efe


In order to consider the confidence levels of the velocity dispersion we 
determine here,
we set $\Delta \chi^2 = 3.5$ for the 68 per cent confidence level of velocity
dispersion, and 7.8 for the 95 per cent confidence level \cite{LMBfirst}, 
because we actually have three adjustable parameters 
($\vsre, L^*$, and the power law between $\sigma_{\re}$ and $L/L^{*}$), even 
though we only vary $\vsre$ and fix the other two parameters. 
The final 95 per cent confidence level is then determined to be 
$\vdm^{*} = 178 {\displaystyle_{-28}^{+29}}$ km s$^{-1}$.

Several issues are relevant to this result.  In the first place, note that the
$L$ vs $\sigma$ relation in the above equation differs from the standard
Faber-Jackson relation \cite{FJ}.  
Note however that this FJ relation is appropriate for
central velocity dispersions.  In addition, it may be there case that the
elliptical galaxies do not form a uniform population, but rather that bright
galaxies and faint galaxies follow separate Faber-Jackson curves (this point
was raised to us by J. Peebles, who has been investigating this issue
\cite{peeblesfuk}). Nakamura and Suto \shortcite{suto} have also pointed
out how the value of the power law relation can differ from 4.
It thus may not be appropriate to enforce this relation on
the bulk sample.  To explore this latter possibility, and because it is
galaxies with the largest velocity dispersions which will dominate in the
analysis of lensing statistics, we considered a subset of our galaxy sample
with $\lv_{\re}$/km s$^{-1}) > 2.2$ (i.e., 23 galaxies)
and rederived the $L$-$\sigma$ relation.
In this case, we find:

\be
\lll = (-6.20 \pm 1.82) + (2.83 \pm 0.79)\lv_{\re}/{\rm km\ s}^{-1})
\label{eq:vre2.2}
\ee
with $\chi^2 = 23.6$ and $\vsre = 155$ km s$^{-1}$.
In this case, we find a somewhat higher value of $\vdm^*
=203$ km s$^{-1}$ as might be expected given the features of this subsample.

A recent result from galaxy-galaxy weak lensing \cite{Fischer} constrains
the velocity dispersion $\sigma_v$ of the average foreground galaxy
to be 150-190 km s$^{-1}$ and this result translates to a $\vdm$ (using
our notation) of
the average foreground galaxy to be
210-266 km s$^{-1}$. It is interesting to compare this result with
the estimate we would derive here.  Based on the
luminosity of the average foreground galaxy given in
Fischer et al.~\shortcite{Fischer}, i.e., $8.7\times 10^9 L_{sun}$ in the
g' band, our formula would predict a
$\vdm$ around 220 km s$^{-1}$, which is in good agreement with the range of
$\vdm$ determined by Fischer.

\subsection{Velocity dispersions at $R = 0.54 \re$}
\label{sec:test}

Another fit between luminosities and velocity dispersions of sample galaxies 
has been calculated at
$R = 0.54\re$. Velocity dispersions at $R = 0.54\re$ are estimated and listed 
in Table~\ref{table:Re}. We still exclude \astrobj{NGC 4251}, 
\astrobj{NGC 7796}, and \astrobj{NGC 4486B},
for the same reasons described in the above section. We find
\be
\lll = (-5.01 \pm 0.55) + (2.28 \pm 0.24)\lv_{0.54\re}/{\rm km\ s}^{-1})
\label{eq:lv54}
\ee
with $\chi^{2}$ = 47 and $\vff^{*} = 159 \pm 15$ km s$^{-1}$ 
(Figure~\ref{fig:0.54re}). As we have discussed in Section~\ref{sec:VD}, if we 
multiply this number by 1.25, then we should get the expected $\vdm^{*}$. 
Therefore, $\vdm^{*} \approx 1.25\vff^{*}$ = 198 km s$^{-1}$ in this case. 
If we consider a subset of our galaxy sample with 
$\lv_{0.54\re}/{\rm km\ s}^{-1}) > 2.2$ (i.e., 28 galaxies), we find
\be
\lll = (-6.43 \pm 1.34) + (2.88 \pm 0.58)\lv_{0.54\re}/{\rm km\ s}^{-1})
\label{eq:lv54_2.2}
\ee
with $\chi^2=23$ and $\vff^*=169$ km s$^{-1}$. In this case, we find 
$\vdm^*=211$ km s$^{-1}$. 

\bfe
\bc\
\epsfxsize=15.0cm
\epsffile{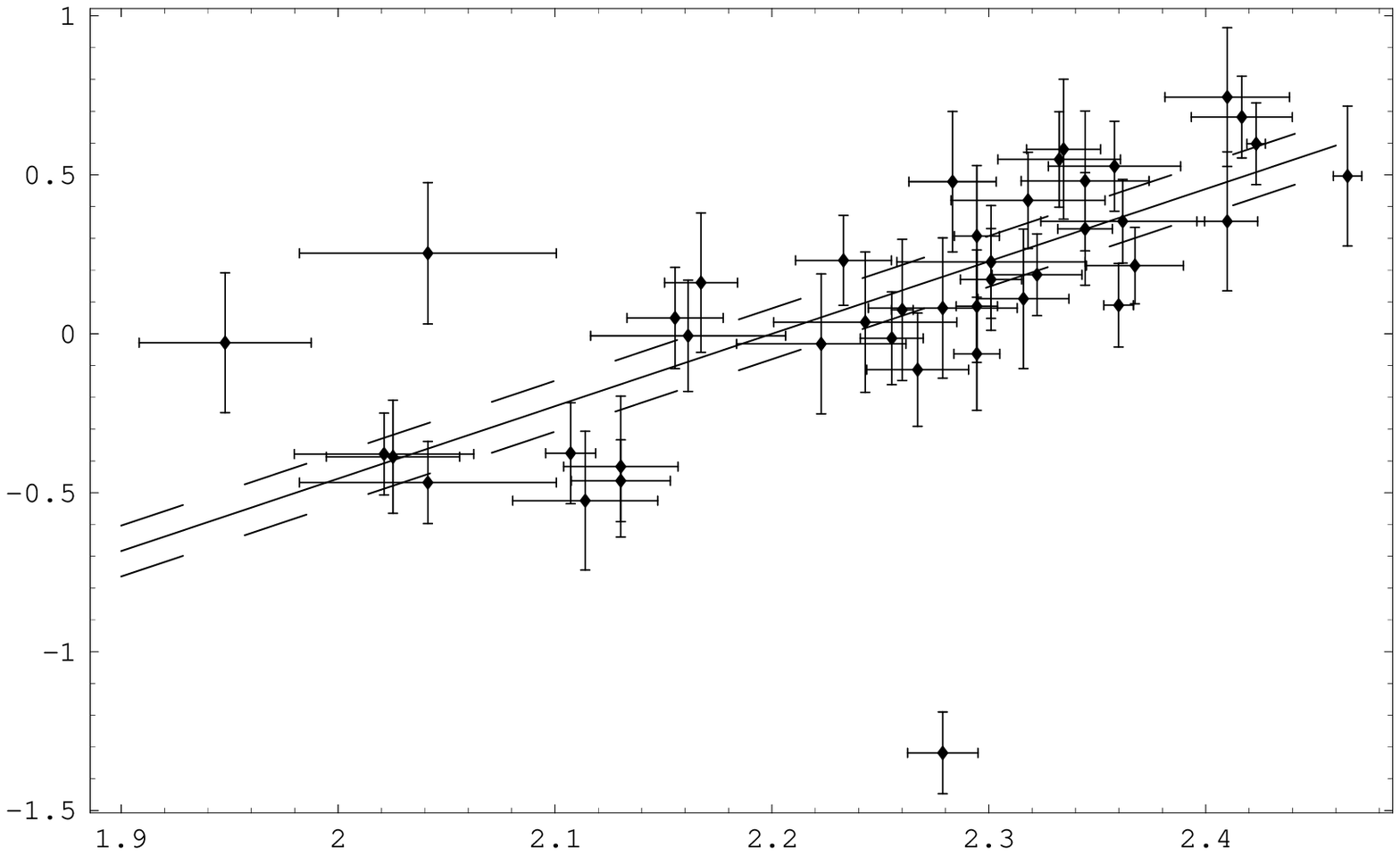}
\ec
\vskip -2.60in
\rotate[l]{\hskip -0pt \large $\lll$}
\vskip 0.3in
{\hskip 3.2in \Large $\chi^{2} = 47$}
\vskip 1.05in
\bc
{\large $\lv_{0.54\re}$/km s$^{-1})$}
\ec
\caption
{Least-square-fit of $\lll$ vs. $\lv_{0.54\re}$/km s$^{-1})$. The solid
line is the best fit, as shown in equation~(\protect\ref{eq:lv54}).
The upper dash line is the best fit plus 0.08, and the lower dash line is the
best fit minus 0.08. These two lines are the 95 per cent confidence limits.}
\label{fig:0.54re}
\efe

We see that the two
slopes in $L$-$\sigma$ relation obtained at different values
of $\re$ are close to each other. Furthermore, 
the inferred value of $\vdm^*$ at $R=0.54 \re$ is somewhat larger than the
value of $\vdm^*$ at $R=\re$. This can be explained as follows.
The above estimates are for the case of a purely finite isothermal
distribution.  If one adds to this distribution some central mass, such as
a large central black hole, this will further increase the central velocity
dispersion, and also change the relationship between $\vdm$ and the
line-of-sight velocity dispersion at $\re$ and half $\re$.  By measuring the
velocity dispersions at both points, however, one can hope to extract out
the central mass contribution and also the isothermal contribution.
(Alternately, it is clear that the velocity dispersion at $\re$ will be less
sensitive to the former contribution, and thus can be used to approximate the
isothermal contribution.) It is important to recognize that the central
mass contribution can affect the velocity dispersion more significantly
than it affects the optical depth for lensing, so that incorporating this
contribution in an effective isothermal sphere velocity dispersion will
lead to an overestimate of the optical depth.

As a quick demonstration, let us see how we can extract out the central mass
contribution. If we simply assume $\nu (x) \propto 1/x^4$ 
due to a point mass at the center of
galaxy, from equations~(\ref{eq:brit}) and~(\ref{eq:model}), we can obtain
\be
\sigma_{\rm los}^2 (R) = \frac{GM}{6R}
\ee
where $M$ is the mass of that point mass. Then, we can set up equations
as follows:
\bea
\sigma_{\rm los}^2(\re) &=& \sigma_{\rm los,g}^2 (\re) + \frac{GM}{6\re} 
\nonumber \\
\sigma_{\rm los}^2(0.54 \re) &=& \sigma_{\rm los,g}^2 (0.54 \re) + 
\frac{GM}{6\cdot (0.54\re)} 
\eea
where $\sigma_{\rm los,g}(R)$ is the isothermal contribution to the 
velocity dispersion of that galaxy as a
function of $R$. As discussed above, we know that $1.31 \sigma_{\rm los,g} (\re)
= 1.25 \sigma_{\rm los,g} (0.54 \re) = \vdm$. Thus, the appropriate
isothermal effective  $\vdm$ can be 
calculated for each galaxy and the value of $\vdm^*$ can be found through a
least-square-fit method. A very quick estimation of the
appropiate $\vdm^*$ from 
equations~(\ref{eq:vre2.2}) and~(\ref{eq:lv54_2.2}), is 
 $\vdm^* \approx$ 190 km s$^{-1}$, with the slope of the $L$ vs. $\vdm$ 
relation being $\approx$
2.8.  It is interesting to note that when values close to these
 are utilized in
an analysis of lensing statistics \cite{chengkrauss} one obtains mean angular
splittings that are comparable to the observed splittings. 

\section{DISCUSSIONS}
\label{sec:conclusion}

First note that both $\mstar$ and $\alpha$, fitted to the Schechter function are
close to the results in Loveday et al.~\shortcite{Loveday}. This is expected, 
because the $\rm B_T$ and $\bj$ bands are close. In fact, we
would expect our $\mstar$ to be slightly fainter than 
${\rm M_{\bj}^{*}}$ (= -19.71 + 5$\log_{10}h$), because the $\bj$ band 
is measured 
at a longer wavelength and most galaxy luminosities are peaker toward the
redder end of the spectrum. On the other hand, 
our $\mstar$ is brighter than ${\rm M_{B_T}^{*}}$ (= -19.37 + 5$\log_{10}h$) 
given in 
EEP. Given the Galactic extinction correction, this is also 
what we would expect. We have
also shown
that the Schechter $\alpha$ of elliptical galaxies is shallower than
that for total galaxies, which is 
consistent with EEP and Loveday et al.~\shortcite{Loveday}.

The ratio of our $\phi^{*}$ to the one in Loveday et al.~\shortcite{Loveday} is
about 9.6 per cent. In comparison with the $\phi^{*}$ in Mobasher, Sharples \& 
Ellis~\shortcite{MSEfirst}, the ratio is about 12 per cent. These two numbers 
indicate the fraction of elliptical galaxies out of total galaxies, and they are 
consistent with the conclusion of Postman \& Geller~\shortcite{PG}, of 12 $\pm$ 2
per cent for this ratio. Our results  here  suggest that the population of
elliptical galaxies is (10.8 $\pm$ 1.2) per cent of all  galaxies.

Regarding our luminosity function determination, 
it is of course still possible that the galaxies 
in our sample set do not form a complete sample set.
Currently,
we calculate $<{\rm\frac{V}{V_{max}}}>$ = 0.45~\cite{Schmidt}. If we had
an incomplete elliptical galaxy sample set, and if we would believe those
`missing' E galaxies are near apparent magnitude 12.19, then we would
expect $<{\rm\frac{V}{V_{max}}}>$ larger than 0.45. Also, according to the theoretical prediction in
Zucca, Pozzetti \& Zamorani~\shortcite{ZPZfirst}, we would then have an 
even a 
dimmer ${\rm M^{*}}$ and a steeper $\alpha$. A fainter ${\rm M^{*}}$
would result an even  smaller $\vdm^*$ than the value calculated in 
Section~\ref{sec:Re} and Section~\ref{sec:test}.

The analysis in Section~\ref{sec:Re}, suggests $\vsre = 136 \pm$ 15
km s$^{-1}$. It implies that $\vdm^{*} = 178 \pm$ 29 km s$^{-1}$. This 
uncertainty is large, because the uncertainties from distances of galaxies in 
the sample set and from $\mstar$ are large. (Also note that the best fit
value if we account for possible central mass concentrations is $\vdm^{*}
\approx 190 $ km s$^{-1}$, which falls within the range included above.)
We also notice that
$\mstar$  is a sensitively coupled to $\vdm^{*}$. For example, if we use
$\mstar$ =  -19.9 + 5$\log_{10}h$~\cite{FT,Kochanek94}, then we will increase
$\vdm^{*}$ by 8\%-12\%. On the other hand, if we use $\mstar$ =
-19.5 + $5\log_{10}h$, then we will decrease $\vdm^{*}$ by 5\%-8\%.
One should be careful about this issue in any strong
gravitational lensing analysis. The completeness of our galaxy sample
discussed in Section~\ref{sec:STY} does not seem to lead to a significant
uncertainty by comparison. For example, different numbers of galaxies
included in our sample produce a variation of 4\%-6\% in
$\vdm$, which is within the values estimated from the uncertainty of $\rm
M^*$ itself. It might be interesting to consider the relation
between $\lll$ and $\lv$/km s$^{-1})$ in other bands (e.g. V or R band, 
if the observation is possible; or a radio survey), to see whether 
including more elliptical galaxies would reduce the intrinsic scatter of this 
power law relation. If so, then the improved value of $\vdm^{*}$ 
and the power law from that survey would be a great help in the study of 
gravitational lensing statistics. Based on our present result,
the optical depth of gravitational lenses would be
expected to be reduced by 43 per cent
compared to the one in Kochanek~\shortcite{Kochanek93}, if all 
other parameters are unchanged. 

In this paper, we treat the anisotropy parameter $\beta$ as a constant.
We expect that our conclusion will not change too much even if $\beta$ is a 
function of radial distance $r$, as long as $|\beta| \leq$ 0.4 
\cite{Kochanek94}
(Figure~\ref{fig:Jaffe} and Figure~\ref{fig:Hernquist}). 

Finally, the theoretical modelling in Section~\ref{sec:VD}, suggests that
$\vvratio$ is strongly dependent on models and parameters within 0.1$\re$ 
(Figure~\ref{fig:Jaffe} and Figure~\ref{fig:Hernquist}).
We also observe that the ratio $\vvratio$ converges at about 0.5$\re$ for all
different parameters. This suggests that we should use observational values 
of $\vvratio$
between 0.4$\re$ and $\re$, if we want to reduce the intrinsic scatter from
theoretical modelling. 
We hope observers will consistently measure velocity 
dispersions of elliptical galaxies at both $\re$ and $0.54 \re$ in order to 
refine our results. As we have discussed in Section~\ref{sec:test}, one
can extract extra velocity dispersions at $\re$ (and $0.54 \re$), and get
an effective value of $\vdm$ for each galaxy. As data improves, we expect that 
the techniques described here will be
useful in further constraining  
input parameters in strong gravitational lensing analysis.

\section*{ACKNOWLEDGMENTS}

We would like to thank Harold Corwin, Deepak Jain,
Peter Kernan, Jon Loveday, Paul Mason, Heather Morrison,
Jim Peebles, Keith Robinson, 
Paul Schechter and the anonymous referee for very helpful input.
This research work has been partially supported by the Industrial Physics
Group in the Physics Department at Case Western Reserve University and by a
grant to the particle astrophysic group at CWRU from the DOE.
This research has also made use of the NASA/IPAC Extragalactic Database (NED)
which is operated by the Jet Propulsion Laboratory, California Institute
of Technology, under contract with the National Aeronautics and Space
Administration.

\newpage
\thispagestyle{empty}

\begin{table}
{\scriptsize
\btr{lcccccc} 
Name & $\re$(\arcsec) & Source & $m(\bt)$ & Distance ($h^{-1}$Mpc) & $\sigma_{\re}$ (km s$^{-1}$) & $\vff$ (km s$^{-1}$)$^c$ \\
\astrobj{NGC 636} & 19.4 & FIH & 12.16 & 15.0 $\pm$ 1.8 & 123 $\pm$ 10 	& 128 $\pm$ 3.4 \\
\astrobj{NGC 720} & 36.1 & BDI & 11.13 & 20.5 $\pm$ 4.4 & 174 $\pm$ 12 	& 197 $\pm$ 4.7 \\
\astrobj{NGC 777} & 34.4 & JS & 12.24 & 44.0 $\pm$ 3.8 & 275 $\pm$ 92 	& 228 $\pm$ 16 \\
\astrobj{NGC 1052} & 33.7 & BDI/DB/FI & 11.33 & 17.2 $\pm$ 3.7 & 176 $\pm$ 3.4 	& 182 $\pm$ 2.1 \\
\astrobj{NGC 1399} & 40.7 & SBS/WF & 10.22 & 14.22 $\pm$ 0.88 & 230 $\pm$ 11 	& 230 $\pm$ 20 \\
\astrobj{NGC 1400} & 29.3 & BZ & 11.87 & 19.9 $\pm$ 1.9 & 170 $\pm$ 25 	& 180 $\pm$ 6.0 \\
\astrobj{NGC 1404} & 23.8 & FIH/WF & 10.88 & 14.22 $\pm$ 0.88 & 209 $\pm$ 8.1 	& 229 $\pm$ 3.6 \\
\astrobj{NGC 1700} & 18.5 & BSG/FIH & 12.00 & 31.4 $\pm$ 4.7 & 195 $\pm$ 7.9 	& 221 $\pm$ 6.4 \\
\astrobj{NGC 2434} & 31.4 & CQ & 11.61 & 19.7 $\pm$ 4.2 & 190 $\pm$ 14 	& 190 $\pm$ 15 \\
\astrobj{NGC 2513} & 32.9 & JS & 12.48 & 46.6 $\pm$ 9.9 & 185 $\pm$ 82 	& 221 $\pm$ 15 \\
\astrobj{NGC 2663} & 54.6 & CQ & 10.23 & 22.4 $\pm$ 4.7 & 220 $\pm$ 14 	& 257 $\pm$ 17  \\
\astrobj{NGC 2778} & 15.7 & JS & 13.21 & 36.3 $\pm$ 7.7 & 71 $\pm$ 26 	& 88.7 $\pm$ 8.1  \\
\astrobj{NGC 3115} & 32.1 & BSG & 9.75 & 10.2 $\pm$ 2.2 & 100 $\pm$ 10 	& 110 $\pm$ 15  \\
\astrobj{NGC 3193} & 26.7 & BSG & 11.73 & 24.6 $\pm$ 3.7 & 145 $\pm$ 25 	& 200 $\pm$ 20 \\
\astrobj{NGC 3377} & 34.4 & BSG/FI & 10.98 & 8.6 $\pm$ 1.3 & 79.2 $\pm$ 4.8  	& 106 $\pm$ 7.5 \\
\astrobj{NGC 3379} & 35.2 & DB/SBS & 10.17 & 8.6 $\pm$ 1.3 & 186 $\pm$ 6.7  	& 197 $\pm$ 4.8 \\
\astrobj{NGC 3557} & 30.0 & FI & 10.79 & 24.0 $\pm$ 5.1 & 208 $\pm$ 15  	& 216 $\pm$ 8.5  \\
\astrobj{NGC 3608} & 33.7 & JS & 11.71 & 19.9 $\pm$ 2.4 & 165 $\pm$ 26  	& 143 $\pm$ 7.3 \\
\astrobj{NGC 3610} & 15.4 & BSG & 11.58 & 21.3 $\pm$ 4.5 & 129 $\pm$ 11  	& 147 $\pm$ 5.7 \\
\astrobj{NGC 3706} & 20.3 & CQ & 11.87 & 30.4 $\pm$ 6.4 & 238 $\pm$ 8.9  	& 257 $\pm$ 8.3 \\
\astrobj{NGC 4251} & 17.7 & BSG & 11.54 & $(10.14 \pm 0.56)^{b}$ & 85 $\pm$ 10  	& 110 $\pm$ 15 \\
\astrobj{NGC 4261} & 36.1 & DB & 11.32 & 27.8 $\pm$ 5.9 & 248 $\pm$ 5.9  	& 292 $\pm$ 4.4 \\
\astrobj{NGC 4278} & 34.4 & DB & 10.96 & 14.7 $\pm$ 2.2 & 178 $\pm$ 12  	& 197 $\pm$ 4.3 \\
\astrobj{NGC 4291} & 17.3 & BSG/JS & 12.32 & 30.3 $\pm$ 3.7 & 187 $\pm$ 18  	& 200 $\pm$ 6.5 \\
\astrobj{NGC 4472} & 104 & SBS/SZ/WF & 9.26 & 13.33 $\pm$ 0.71 & 198 $\pm$ 9.2  	& 261 $\pm$ 14 \\
\astrobj{NGC 4486B} & (3.07)$^{a}$ & BSG & 14.26 & 13.33 $\pm$ 0.71 & 167 $\pm$ 8.3  	& 190 $\pm$ 7.1 \\
\astrobj{NGC 4486} & 94.9 & WF & 9.47 & 13.33 $\pm$ 0.71 & 238 $\pm$ 12  	& $(265 \pm 2.6)^{d}$ \\
\astrobj{NGC 4494} & 48.7 & JS & 10.61 & 7.0 $\pm$ 1.5 & 127 $\pm$ 21  	& 135 $\pm$ 8.2 \\
\astrobj{NGC 4564} & 19.8 & BSG & 11.91 & 13.33 $\pm$ 0.71 & 80 $\pm$ 10  	& 105 $\pm$ 10 \\
\astrobj{NGC 4621} & 40.5 & Bender & 10.5 & 13.33 $\pm$ 0.71 & 140 $\pm$ 40  	& 210 $\pm$ 10 \\
\astrobj{NGC 4660} & 12.2 & BSG & 12.12 & 13.33 $\pm$ 0.71 & 110 $\pm$ 10  	& 135 $\pm$ 7.1 \\
\astrobj{NGC 5018} & 22.8 & CQ & 11.34 & 29.8 $\pm$ 3.1 & 210 $\pm$ 7.1  	& 215 $\pm$ 14 \\
\astrobj{NGC 5576} & 18.1 & BSG & 11.71 & 16.5 $\pm$ 2.5 & 100 $\pm$ 20  	& 185 $\pm$ 10 \\
\astrobj{NGC 5812} & 25.5 & BZ & 11.87 & 21.1 $\pm$ 4.5 & 173 $\pm$ 117  	& 175 $\pm$ 17 \\
\astrobj{NGC 6411} & 28.6 & JS & 12.64 & 27.8 $\pm$ 5.9 & 333 $\pm$ 120  	& 167 $\pm$ 15 \\
\astrobj{NGC 7144} & 32.1 & SZ & 11.68 & 18.4 $\pm$ 2.7 & 155 $\pm$ 14  	& 145 $\pm$ 15 \\
\astrobj{NGC 7454} & 25.0 & BSG & 12.7 & 16.2 $\pm$ 3.4 & 97.7 $\pm$ 8.3  	& 130 $\pm$ 10 \\
\astrobj{NGC 7507} & 30.7 & BZ & 11.28 & 17.5 $\pm$ 3.7 & 180 $\pm$ 15  	& 207 $\pm$ 10 \\
\astrobj{NGC 7626} & 38.6 & JS & 12.06 & 35.8 $\pm$ 3.8 & 166 $\pm$ 86  	& 208 $\pm$ 17 \\
\astrobj{NGC 7785} & 23.3 & BSG & 12.41 & 45.0 $\pm$ 9.6 & 210 $\pm$ 30  	& 192 $\pm$ 8.9 \\
\astrobj{NGC 7796} & 21.2 & BZ & 12.39 & (32.9 $\pm 0.24)^{b}$ & 225 $\pm$ 37  	& 233 $\pm$ 12 \\
\astrobj{IC 179} & 16.5 & JS & 12.98 & 44.0 $\pm$ 3.8 & 124 $\pm$ 21   	& 171 $\pm$ 8.7 \\
\etr
}\\
$^{a}$ This value is from BSG.\\
$^{b}$ The local peculiar velocity contributes to the distance estimate in  RC3.\\
$^c$ Reference WF~\protect\cite{WF} is not used.\\
$^{d}$ This value is from DB.
\vskip 6pt
\caption{The galaxy sample used in deriving the velocity 
dispersion-luminosity relation. The names of galaxies are listed in the first 
column. Their corresponding effective radii, observational sources, apparent 
magnitudes, distances, velocity dispersions at effective radii, and velocity 
dispersions at 0.54 effective radii
are listed in columns 2 to 7, respectively.} 
\label{table:Re}
\end{table}

\end{document}